\documentstyle[12pt]{article}

\def\overlay#1#2{\ifmmode \setbox 0=\hbox {$#1$}\setbox 1=\hbox to\wd 0{\hss
$#2$\hss }\else \setbox 0=\hbox {#1}\setbox 1=\hbox to\wd 0{\hss #2\hss }\fi
#1\hskip -\wd 0\box 1}

\catcode`@=11
\newcount\@tempcntc
\def\@citex[#1]#2{\if@filesw\immediate\write\@auxout{\string\citation{#2}}\fi
  \@tempcnta\z@\@tempcntb\m@ne\def\@citea{}\@cite{\@for\@citeb:=#2\do
    {\@ifundefined
       {b@\@citeb}{\@citeo\@tempcntb\m@ne\@citea\def\@citea{,}{\bf ?}\@warning
       {Citation `\@citeb' on page \thepage \space undefined}}%
    {\setbox\z@\hbox{\global\@tempcntc0\csname b@\@citeb\endcsname\relax}%
     \ifnum\@tempcntc=\z@ \@citeo\@tempcntb\m@ne
       \@citea\def\@citea{,}\hbox{\csname b@\@citeb\endcsname}%
     \else
      \advance\@tempcntb\@ne
      \ifnum\@tempcntb=\@tempcntc
      \else\advance\@tempcntb\m@ne\@citeo
      \@tempcnta\@tempcntc\@tempcntb\@tempcntc\fi\fi}}\@citeo}{#1}}
\def\@citeo{\ifnum\@tempcnta>\@tempcntb\else\@citea\def\@citea{,}%
  \ifnum\@tempcnta=\@tempcntb\the\@tempcnta\else
   {\advance\@tempcnta\@ne\ifnum\@tempcnta=\@tempcntb \else \def\@citea{--}\fi
    \advance\@tempcnta\m@ne\the\@tempcnta\@citea\the\@tempcntb}\fi\fi}
\catcode`@=12
\begin{document}

\font\fortssbx=cmssbx10 scaled \magstep2
\hbox to \hsize{
\hskip.5in \raise.1in\hbox{\fortssbx Rutherford Appleton Laboratory}
\hfill\vtop{\hbox{\bf RAL-94-017}
                \hbox{January 1994}}}

\vspace{.5in}

\begin{center}
{\large\bf SOLAR NEUTRINO OSCILLATIONS}\\[.2in]
R.J.N.~Phillips\\[.1in]
{\it
Rutherford Appleton Laboratory, Chilton, Didcot, Oxon OX11 0QX, England}
\end{center}

\vspace{.5in}
\begin{abstract}
This is an invited review talk, presented at the International
Conference on Non-Accelerator Particle Physics (ICNAPP-94),
Bangalore, India, 2-9 January 1994.
\end{abstract}

%\thispagestyle{empty}

%\newpage

{\bf 1. Introduction}.

Standard Solar Models (SSM) predict the $\nu_e$ flux of Fig.1 with some
uncertainties\cite{bp,tc}. Measurements by capture in $^{37}Cl$ \cite{cl},
$\nu-e$ scattering\cite{kam} and capture in $^{71}Ga$ \cite{sage,gallex},
with differing $E_{\nu}$ thresholds, find three different deficits:
$$
\begin{array}{llll}
Detection & Threshold & Observation/SSM[1] & Observation/SSM[2] \\
\nu -e    & 7.5 MeV  & 0.51\pm .07\pm .07   & 0.66\pm .09\pm .16 \\
^{37}Cl   & 0.81 MeV & 0.29\pm .03\pm .04   & 0.36\pm .04\pm .08 \\
^{71}Ga   & 0.24 MeV & 0.62\pm .10\pm .03   & 0.67\pm .11\pm .04 \\
\end{array}
$$
where the first error is experimental, the second is from the SSM.
These numbers suggest a differential suppression, with the top and bottom
of the accessible range less suppressed than the middle. They pose the
Solar Neutrino Problem 1994.

Re-tuning the solar model gives no easy solution\cite{nssm}. A lower
central temperature would suppress $^8B$ production and the $\nu-e$
rate, but to explain $^{37}Cl$ rates the $^7Be$ line must then
be obliterated - a bit unlikely given that $^8B$ is made from $^7Be$.

Neutrino oscillations offer several possible explanations, that I
briefly compare.

{\bf 2. Long Wavelength Vacuum Oscillations (LWVO)}.

Suppose the weak eigenstate $\nu_e$, emitted by $\beta$-decays in the
Sun, is actually a superposition of two mass eigenstates: $\nu_e = \nu_1
cos\theta - \nu_2sin\theta$ with $\nu_{\mu} = \nu_1sin\theta +
\nu_2cos\theta$.  The mass eigenstates propagate
independently with time t, each picking up a different phase factor
$exp(-im_i^2t/2E)$, so that after a distance $L=ct$ the projection back
onto $\nu_e$ becomes $A(\nu_e \to \nu_e) = [\cos^2\theta exp(-im_1^2L/2E)
+ \sin^2\theta exp(-im_2^2L/2E)]$.  The probability that this evolved state
can interact like $\nu_e$ is then
$$
P(\nu_e \to \nu_e) = |A|^2 = 1-sin^2(2\theta)sin^2(\delta m^2 L/4E) .
$$
where $\delta m^2 = m_2^2 - m_1^2$.  Figure 2 illustrates this
oscillatory probability.  For $E/\delta m^2 >> 1$ there is negligible
effect; for values $\sim 0.1-1$ there are resolvable oscillations;
for values $<< 0.1$ the oscillations are averaged in practice, either
by source/detector size or by energy resolution. Averaged 2-neutrino
oscillations suppress by at most 1/2, but n-neutrino mixing can
give 1/n; however this suppression is flat and therefore
unsuited to the 1994 solar problem.  On the other hand, resolved
oscillations provide a strongly varying suppression.  Try overlaying
Figs.1 and 2 (they have the same horizontal log scale).  If we tune
$\delta m^2$ such that the first minimum falls around a few MeV, the
Kamiokande $\nu -e$ rate will be somewhat suppressed; if we fine-tune
to put the $860 KeV~ ^7Be$ line in a minimum, the $^{37}Cl$ rate
will be somewhat more suppressed; meanwhile the $^{71}Ga$ rate suffers
less, since its dominant pp neutrinos encounter only average suppression.
We clearly have the makings of one or more solutions here, with these
``just-so" oscillations\cite{lwvo}.  Figure 3 shows typical recent fits
\cite{kras} in the $(sin^22\theta, \delta m^2$) parameter plane; the
disconnected regions put the $^7Be$ line in different minima of P.  Note
that for $\nu_e - \nu_{\mu}$ mixing, $\nu_{\mu}-e$ scattering contributes
a bit to the Kamiokande signal and helps to explain why it is less
suppressed than $^{37}Cl$; $\nu_e - \nu_x$ sterile flavour mixing
lacks this help and is harder to fit.

Two special features arise from LWVO resolved oscillation
patterns \cite{lwvo}.\\
i) There is an oscillatory modulation on the shape of the high-
energy $^8B$ spectrum contribution. The shape (though not the magnitude)
of this SSM component is model-independent; the modulation would  be
detectable at SNO\cite{wark} and Super-Kamiokande.  See Figs. 8,9 at
the end.\\
ii) The $^7Be$ line with fixed $E_{\nu}$ has oscillatory strength,
because the Earth-Sun distance $L$ has small seasonal variations.
This line strength could be measured directly by Borexino\cite{borexino};
the effect is diluted in $^{37}Cl$ and $^{71}Ga$ signals. \\
(Present $ \nu-e$, $Cl$ and $Ga$ data constrain (i) and (ii) rather
weakly).

But mixing $\nu_e$ with $\nu_{\mu}$ or $\nu_{\tau}$ affects $\nu_e$ spectra
from supernovae; SN1987A data may disfavour large $sin^2(2 \theta ) >
0.7-0.9$ \cite{ssb}, including solutions like Fig.3a.

{\bf 3. Oscillations in matter}.

Coherent forward scattering in matter generates a refractive index and
affects propagation\cite{wolf}.  Z-exchange processes are the same for
$\nu_e$, $\nu_{\mu}$, $\nu_{\tau}$, generating a common phase that can
be ignored, but W-exchange contributes only to $\nu_e-e$ scattering and
significantly changes the propagation equation:
$$
4iE{d\over dt}\pmatrix{\nu_e\cr\nu_{\alpha}\cr}
=\pmatrix{m_1^2+m_2^2-\delta m^2cos2\theta+4\sqrt 2G_F\rho_e E
          &\delta m^2sin2\theta\cr
           \delta m^2sin2\theta
          &m_1^2+m_2^2+\delta m^2cos2\theta \cr}
 \pmatrix{\nu_e\cr\nu_{\alpha}\cr}
$$
where $\theta$ is the vacuum mixing angle, $\rho_e$ is the electron
number density and $\nu_{\alpha}= \nu_{\mu}$ or $\nu_{\tau}$ (for
sterile $\nu_x$ see later).  Diagonalizing the propagation matrix
above, we find that the mixing angle in matter $\theta_m$ depends
on $\rho_e E$:
$$
tan2\theta_m=\frac{tan2\theta}
                    {1-(2\sqrt 2G_F\rho_e E)/(\delta m^2cos2\theta)}.
$$
If $\delta m^2 > 0$ the mixing is enhanced; it becomes maximal
($\theta_m = \pi/4$) where the denominator vanishes - sometimes called
a resonance.  As neutrinos travel out from the solar core, the mixing
angle $\theta_m$ and the matter-propagation eigenstates $\nu_{1m},
\nu_{2m}$ change continuously.

This gives a possibility for  efficient $\nu_e \to \nu_{\mu}(\nu_{\tau})$
conversion via adiabatic level crossing (the MSW effect\cite{msw}).
Suppose that $\rho_e E$ is far above the resonance value at the point of
$\nu_e$ creation in the solar core (i.e. $\theta_m \sim \pi/2$); then
$\nu_e \simeq \nu_{2m}$ here. If subsequent propagation is adiabatic,
the local eigenstate components are essentially preserved: $\nu_{jm}
\to \nu_{jm} (j=1,2)$.  Emerging from the Sun, the dominant $\nu_{2m}$
component becomes the vacuum mass eigenstate $\nu_2=-\nu_esin\theta
+\nu_{\mu}cos\theta \simeq \nu_{\mu}$ if the vacuum mixing angle
$\theta$ is small; thus initial $\nu_e$ ends up as mostly $\nu_{\mu}$.
Fig.4 shows how the two eigenvalues $m_j^2$ of the propagation matrix
behave versus $\rho_e$; solid lines give the case of no mixing,
$\theta = 0$; dashed lines show how the eigenvalues cross over
when mixing is present.  If $\rho_e$ changes slowly enough for the
mixing to act (adiabatically), the physical state follows the
full eigenstates (dashed lines); but if $\rho$ changes too suddenly,
the physical state follows the unmixed eigenstates (solid lines).

There are 2 conditions for adiabatic level crossing.\\
(i) Central density is above the resonance value:
$$
E(MeV)/\delta m^2(eV^2) > cos2\theta/[2\sqrt 2G_F\rho_e(max)]
                        \simeq 10^5,
$$
(ii) Density changes slowly enough near the resonance for adiabaticity:
from the Landau-Zener approximation we obtain\cite{parke}
$$
E(MeV)/\delta m^2(eV^2) << sin^22\theta/[\rho_e^{-1}d\rho_e/dR]cos2\theta
                        \simeq 3 \times 10^8 sin^22\theta.
$$
The detailed consequences require big calculations, but we need no
computer to see the main features.
The MSW effect gives a bathtub-shaped suppression factor; see Fig.5.
The steep left-hand end is determined by the resonance-crossing
condition (i); the sloping right-hand end covers the range where
adiabaticity breaks down, determined by condition (ii). We can choose
almost any bathtub we please, versus energy $E_{\nu}$, by selecting
$\delta m^2 cos2\theta$ to get the left-hand end and $tan^22\theta$
to get the length (and also the depth) of the bathtub. However, condition
(ii) excludes MSW effects in the LWVO region. Notice also that
efficient $\nu_e \to \nu_{\mu}$ conversion does not require big
vacuum mixing $\theta$; on the contrary, the best conversion is with
small $\theta$.

The MSW bathtub offers an immediate explanation of the
apparent differential suppression of the $\nu_e$ spectrum: let the
sloping end lie across the $\nu-e$ scattering range $E_{\nu}>7.5 MeV$
(moderate $\nu-e$ suppression); let the flat bottom lie across
the rest of the $^{37}Cl$ capture range $E_{\nu} > 0.9 MeV$
(more $^{37}Cl$ suppression); let the steep end fall near
$E_{\nu} \sim 0.2 MeV$ at the top of the $pp$ spectrum
contribution (less $^{71}Ga$ suppression).  This simple prescription
leads to the best MSW solution: $\delta m^2 \sim 10^{-5} eV^2$ with
$sin^22\theta \sim 10^{-2}$,  Fig.6a shows a typical recent
fit\cite{hala}.  There is also a  large-$\theta$ region, not really a
good solution but a local $\chi^2$ minimum, where the bathtub is much
shallower and wider.

Matter effects can also arise in the Earth. They are not MSW (no
chance of adiabatic level crossing), just amplified vacuum oscillations
through which $\nu_{\mu}$ can convert back to $\nu_e$ when the sun
is below the horizon, giving day/night and summer/winter asymmetries
in counting rates. There are 2 conditions for big Earth effects.\\
(i) Near-resonant amplification in the Earth ($\tan 2\theta_m$ large):
$$
\delta m^2(eV^2)cos2\theta /E(MeV) \sim 2\sqrt 2G_F \rho_e
                                   \simeq 3\times 10^{-7} ,
$$
assuming rock density $\sim 4 gm/cm^3$.\\
(ii) Matter oscillation wavelength ($\lambda = 4\pi E/\delta m_m^2$)
less than Earth diameter ($10^7 m$). At resonance the matter-eigenvalue
difference is $\delta m_m^2 = \delta m^2 sin2\theta$, giving
$$
\delta m^2(eV^2)sin2\theta/E(MeV) > 2.5\times 10^{-7}.
$$
Both these conditions can be approached or satisfied in a small region
of $(\delta m^2, sin^22\theta)$ for given $E$, but $\theta$
cannot be very small.  For the Kamiokande $\nu-e$ range, $E\sim10$ MeV,
a region near the MSW large-$\theta$ solution is sensitive to Earth
effects; the absence of a day/night asymmetry\cite{kam} excludes this
region (labelled "excluded 90\% C.L." in Fig.6). Future experiments will
enlarge this region of sensitivity.

Similar things can happen for $\nu_e$ mixing with sterile $\nu_x$, but
now Z-exchange no longer drops out; coherent $\nu_e-e$ and $\nu_e-p$
Z-exchanges cancel and the net effect is to replace $\rho_e$
above by $(\rho_e -{1\over 2}\rho_n)$ where $\rho_n$ is the neutron
number density\cite{bdppw}.  The critical parameters change a bit
and the large-$\theta$ solution vanishes (Fig.6b).

Three- or four-flavour neutrino mixing offers more complicated
possibilities, with more free parameters, that we do not need yet and
shall not discuss today.

{\bf 4. Exotic neutral current effects.}

If there are new neutral-current interactions, such as $\nu_ed
\to \nu_ed,\nu_{\tau}d$ flavour-conserving or flavour-flipping
scattering via R-parity-violating b-squark exchanges\cite{rpv},
new terms will appear in the matter-propagation matrix.  In the
most general case with diagonal and off-diagonal contributions
from scattering on $e,p,n$ distributions in matter, this matrix
can be put in the form
$$
\pmatrix{m_1^2+m_2^2-\delta m^2cos2\theta + 4\sqrt 2G_F\rho_e E &
          \delta m^2sin2\theta + \epsilon 4\sqrt 2G_F\rho_eE \cr
          \delta m^2sin2\theta + \epsilon 4\sqrt 2G_F\rho_eE    &
   m_1^2+m_2^2+\delta m^2cos2\theta +\epsilon'4\sqrt 2G_F\rho_eE \cr}
$$
in the approximation $\rho_n \simeq \rho_e$, with just two
constant parameters $\epsilon$ and $\epsilon'$ describing
the new physics in units of the standard matter effect.  If
$\epsilon \ne 0$, we have mixing and oscillations even in the
absence of vacuum mixing ($\delta m^2 sin2\theta = 0$)\cite{wolf}.

These new terms modify the previous MSW solutions. Fig.7 compares
$\nu_e-\nu_{\tau}$  solutions in the cases $\epsilon=0$, $\epsilon=0.04$,
$\epsilon=-0.04$ (with $\epsilon'=0$)\cite{fogli}.  Adding this small
exotic mixing scarcely affects the large-$\theta$ solution but
distorts or even splits the small-$\theta$ solution.

{\bf 5. Outlook.}

The different two-flavour-mixing scenarios can be distinguished (or
rejected) by future measurements of the $^8B$ spectrum modulation
(SNO, Super-Kamiokande, see Figs.8,9), the $^7Be$ 0.86 MeV line
contribution (Borexino), and possible day/night effects (SNO,
Super-Kamiokande, ICARUS):
$$
\begin{array}{llll}
Measurement & LWVO & MSW small-\theta & MSW large-\theta \\
^8B \; modulation &  yes  & yes   & none   \\
^7Be\; line       & seasonal & small & medium \\
day/night \; effects &  none & small & medium \\
\end{array}
$$
Furthermore, the charged-current/neutral-current event ratio [SNO,
Borex, ICARUS] will distinguish whether the neutrino flavour mixed
with $\nu_e$ is active ($\nu_{\mu}, \nu_{\tau}$) or sterile
($\nu_x$). The problem will become much more clearly defined.

\section*{Figure captions}
\begin{enumerate}
\item[Fig.~1:] Solar neutrino spectrum in the SSM [1].
\item[Fig.~2:] Typical oscillation factor.
\item[Fig.~3:] Typical LWVO solutions [8].
\item[Fig.~4:] Eigenvalues of propagation matrix versus density $\rho_e$.
\item[Fig.~5:] The MSW bathtub; suppression factor $P(\nu_e \to \nu_e)$
versus $E/\delta m^2$.
\item[Fig.~6:] MSW solutions [16].
\item[Fig.~7:] Typical exotic neutral current effects on MSW solutions [19].
\item[Fig.~8:] Modulation factors for $^8B$ neutrino spectrum.
\item[Fig.~9:] Electron spectra due to $^8B$ neutrinos [16].
\end{enumerate}
\end{document}